\documentclass[aps,prd,nofootinbib,preprint,tightenlines,longbibliography,showpacs,showkeys]{revtex4-1}

\usepackage{tensor}
\usepackage{amsmath}
\usepackage{amssymb}
\usepackage{bbm}
\usepackage{mathrsfs}
\usepackage{mathtools}
\usepackage{verbatim}
\usepackage[T1]{fontenc}
\usepackage{url}

\begin{document}

\title{Shape Dynamical Loop Gravity from a Conformal Immirzi Parameter}

\author{Patrick J Wong}
\email[pjw@thp.uni-koeln.de]{}

\affiliation{Institute for Theoretical Physics\\University of Cologne}

\date{\today}
\begin{abstract}
The Immirzi parameter of loop quantum gravity is a one parameter ambiguity of the theory whose precise interpretation is not universally agreed upon. It is an inherent characteristic of the quantum theory as it appears in the spectra of geometric operators, despite being irrelevant at the classical level. The parameter's appearance in the area and volume spectra to the same power as the Planck area suggest that it plays a role in determining the fundamental length scale of space. In fact, a consistent interpretation is that it represents a constant rescaling of the kinematical spatial geometry. 
An interesting realization is that promoting the Immirzi parameter to be a general conformal transformation leads to a system which can be identified as analogous to the linking theory of shape dynamics. A three-dimensional gravitational gauge connection is then constructed within the linking theory in a manner analogous to loop quantum gravity, thereby facilitating the application of the established procedure of loop quantization. 
\end{abstract}

\pacs{4.20.Fy, 4.60.Pp}
\keywords{Immirzi parameter, shape dynamics}

\maketitle

\section{Introduction}

One of the open problems in the formalism of loop quantum gravity is the nature of the $\beta$ parameter~\cite{cianfrani,livine2009,sahlmann2012,rovelli1998,wongthesis}. This $\beta$ is the source of a one-parameter ambiguity in the quantization of the theory which arises from the fact that $\forall \beta \in \mathbbm{R}\setminus\{0\} = GL(1,\mathbbm{R})$ the classical phase spaces are equivalent, but, from e.g. the eigenvalues of the area operator~\cite{ashtekar1997,frittelli1996}, it is seen that the quantum theory explicitly depends on this quantity. In addition to expected dependence on the Planck length $\ell_p = \sqrt{\varkappa \hbar}$, the quanta of areas and volumes also depend on $\beta$. 
This parameter was originally introduced by Barbero~\cite{barbero1995} and Immirzi~\cite{immirzi1997a,immirzi1997b} as an attempt to create $\mathbbm{R}$-valued versions of the complex variables used by Ashtekar~\cite{ashtekar1986,ashtekar1987} in a Hamiltonian gauge theory of gravity. The original variables could be recovered from choosing $\beta = \pm i$. Although the gauge connection of the Ashtekar formalism was of the full $SL(2,\mathbbm{C})$ Lorentz group, a quantization of this theory was problematic due to 1) the complex nature of the variables forced the implementation of `reality conditions' which could not be quantized easily, and 2) it was (and still is) unknown how to quantize theories with a non-compact gauge group such as $SL(2,\mathbbm{C})$ since only for a compact group $G$ is it possible to construct an orthonormal basis for the Hilbert space $L_2(G,\mathrm{d}\mu)$ with Haar measure $\mathrm{d}\mu$ by the Peter-Weyl theorem. However, the choice of $\beta \in GL(1,\mathbbm{R})$ for Lorentzian spacetime is considered to be aesthetically undesirable~\cite{thiemann} since this choice means that the connection is not the pull-back of a spacetime connection~\cite{samuel2000}. 
The reduction of the gauge group from $SL(2,\mathbbm{C}) \to SU(2)$ also requires choosing the `time gauge' for the tetrads, meaning that internal boosts are no longer a gauge degree of freedom as they have been gauge-fixed to zero. 

Typically, $\beta$ is assumed to be another fundamental constant which needs to be fixed by some experimental or observational measurement. The standard method of fixing this parameter is through calculation of the Bekenstein-Hawking black hole entropy. This fixes the ambiguous parameter to the value $\beta = 0.237~532~957~965~92\ldots$~\cite{meissner2004}. This is the standard result presented in the literature, although more recently there have been calculations which manage to recover the correct Bekenstein-Hawking formula without fixing $\beta$~\cite{ghosh2011}. Although this parameter is accepted as a neccesary part of the theory, it is not agreed upon as to what the physical significance of this parameter is. The appearance of $\beta$ in the spectrum of the geometric operators suggests the interpretation that $\beta$ may have some fundamental relation to the theory's definition of length scales. In particular, it appears that the fundamental length scale of the theory is $\tilde{\ell}_p = \sqrt{\beta} \ell_p$ rather than simply $\ell_p$ since area eigenvalues are proportional to $\beta \ell_p^2$ and volume eigenvalues are proportional to $\beta^{3/2}\ell_p^3$~\cite{ashtekar1998}. A previous exploration of this fact revealed that this rescaling interpretation holds at the kinematical level, but not dynamically \cite{garay2002}.

This article further explores the interpretation of $\beta$ as a scaling factor of the spatial 3-geometry and how its generalization to a non-constant conformal parameter facilitates a connection between loop quantum gravity and the theory of shape dynamics. Section~\ref{S:immirziscale} gives a brief account of how the presence of $\beta$ arises from a rescaling of the spatial geometry, Section~\ref{S:sd} reviews the construction of shape dynamics from general relativity via a linking theory, Section~\ref{S:connectionsd} shows how a gauge connection may be constructed within this linking theory, Section~\ref{S:qsdlg} sketches a possible quantization procedure for the resulting shape dynamical loop gravity, and Section~\ref{S:cc} concludes with some comments on the conceptual compatibility between loop quantum gravity and shape dynamics. The treatment of shape dynamics in this paper concentrates on the asymptotically flat construction. The spatially compact case of shape dynamics is reviewed in Appendix~\ref{S:compact}. In the notation of this paper indices $\{a,b,c,\ldots\}$ label spatial coordinates on $\Sigma$ and indices $\{i,j,k,\ldots\}$ are $\mathfrak{su}(2)$ Lie algebra labels. The functional exterior derivative is denoted by $\mathfrak{d}$ and all `Tr's are performed over $\mathfrak{su}(2)$. 

\section{$\beta$ as a scale factor}\label{S:immirziscale}
The starting point of the theory is general relativity in the canonical picture (`geometrodynamics') wherein the 4d pseudo-Riemannian spacetime manifold $(M , \boldsymbol{g})$ is assumed to be globally hyperbolic and is decomposed into 3d Riemannian spatial foliations $(\Sigma , \boldsymbol{q})$ through the decomposition $M = \Sigma\times\mathbbm{R}$. The conjugate momentum to $q_{ab}$ is given by $p^{ab} = \frac{\sqrt{q}}{2\varkappa}(K^{ab} - q^{ab} K)$ where $K^{ab}$ is the extrinsic curvature tensor. 

The symplectic structure of general relativity in the Ashtekar-Barbero variables is
\begin{equation}
	\Omega_{\mathrm{AB}} = \frac1\beta \int_\Sigma \mathrm{d}^3x \mathfrak{d} \tensor{A}{^i_a} \wedge \mathfrak{d} \tensor{E}{_i^a}
\end{equation}
where the canonical variables $\boldsymbol{A} \equiv \tensor{A}{^i_a}(x) \tau_i \otimes \mathrm{d} x^a$ and $\boldsymbol{E} \equiv \tensor{E}{_j^b}(x) \tau^{*j} \otimes \partial_b$ are a $\mathfrak{su}(2)$-valued connection 1-form and a $\mathfrak{su}(2)^*$-valued vector density respectively. 
These variables have dimension $[\tensor{\boldsymbol{A}}{}] = ML^{-2}$ and $[\tensor{\boldsymbol{E}}{}] = 1$ in units where $c=1$. 
Decomposing the metric on $\Sigma$ into sections of the frame bundle as $q_{ab} = \eta_{ij} \tensor{e}{^i_a} \tensor{e}{^j_b}$, the canonical momentum is related to the metric through $\tensor{E}{_i^a} = \tensor{\epsilon}{_{ijk}} \tensor{\epsilon}{^{abc}} \tensor{e}{^j_b} \tensor{e}{^k_c}$, yielding its physical interpretation as the vector density adjoint of an area element. 
The configuration variable $\tensor{A}{^i_a}$ is defined as $\varkappa \tensor{A}{^i_a} = \tensor{\Gamma}{^i_a} + \beta \tensor{K}{^i_a}$ where $\tensor{\Gamma}{^i_a}$ is the spin connection and $\tensor{K}{^i_a}$ is related to the extrinsic curvature tensor through $\tensor{K}{^i_a} = \eta^{ij} \tensor{e}{_j^b} K_{ab}$. The Hamiltonian vector fields are given by 
\begin{equation}
	X_F = \beta \int_\Sigma \mathrm{d}^3x \left[ \frac{\delta F}{\delta \tensor{A}{^i_a}} \frac{\delta}{\delta \tensor{E}{_i^a}} - \frac{\delta F}{\delta \tensor{E}{_i^a}} \frac{\delta}{\delta \tensor{A}{^i_a}} \right] 
\end{equation}
which leads to a Poisson structure of $\Omega(X_A,X_E) = \left\{ \tensor{A}{^i_a}(x) , \tensor{E}{_j^b}(y) \right\} = \beta \delta^i_j \delta_a^b \delta^3(x,y)$. The existence of a Gau{\ss} law constraint $\mathrm{D} \boldsymbol{E}^\flat = \mathrm{d} \boldsymbol{E}^\flat + \varkappa \boldsymbol{A} \wedge \boldsymbol{E}^\flat = 0$ allows for the interpretation of $\boldsymbol{A}$ as a gauge connection. The phase space of geometrodynamics is equivalent to this phase space modulo $SU(2)$ gauge transformations. 

The appearance of $\beta$ in the $\mathfrak{su}(2)$ connection can be explained as a constant Weyl transformation on the 3-geometry as follows:
\begin{subequations}
\begin{align}
	\Theta	&=	\int_\Sigma \mathrm{d}^3x p^{ab} \mathfrak{d} q_{ab}	\\
	&=	\frac{1}{\varkappa} \int_\Sigma \mathrm{d}^3x \tensor{E}{_i^a} \mathfrak{d} \tensor{K}{^i_a}	\\
	&=	\frac{1}{\varkappa} \int_\Sigma \mathrm{d}^3x \frac{\tensor{E}{_i^a}}{\beta} \mathfrak{d} \left( \beta \tensor{K}{^i_a} \right)	\\
	&=	\frac{1}{\varkappa} \int_\Sigma \mathrm{d}^3x \frac{\tensor{E}{_i^a}}{\beta} \mathfrak{d} \left( \beta \tensor{K}{^i_a} \right) + \frac{1}{\varkappa} \int_\Sigma \mathrm{d}^3x \frac{\tensor{E}{_i^a}}{\beta} \mathfrak{d} \tensor{\Gamma}{^i_a}	\\
	&=	\frac{1}{\beta\varkappa} \int_\Sigma \mathrm{d}^3x \tensor{E}{_i^a} \mathfrak{d} \left( \tensor{\Gamma}{^i_a} + \beta \tensor{K}{^i_a} \right)	\\
	&=	\frac{1}{\beta} \int_\Sigma \mathrm{d}^3x \tensor{E}{_i^a} \mathfrak{d} \tensor{A}{^i_a}
\end{align}
\label{E:betatrans}
\end{subequations}
Note that $\int \mathrm{d}^3x \tensor{E}{_i^a} \mathfrak{d} \tensor{\Gamma}{^i_a}$ is a boundary term for $\Sigma$ asymptotically flat and $\Gamma[E] = \Gamma[E/\beta]$~\cite{thiemann}. 
In the context presented here, the Immirzi parameter is tied to the 3-geometry and the Hamiltonian picture; it is only introduced \textit{after} the $3+1$ decomposition has been performed. This provides an interpretation that $\beta$ is related to the slicing of spactime into spatial foliations. 

The approach and interpretation here is independent of Lagrangian appearances of the Immirzi parameter, e.g. Refs.~\cite{baekler2011,castillofelisola2015,mercuri2009}, where $\beta$ is often interpreted to be a topological parameter. In covariant loop quantum gravity, $\beta$ explicitly appears in the $Y_\gamma$ map\footnote{In the covariant loop quantum gravity literature, the Immirzi parameter is denoted by $\gamma$.} which plays an integral role in determining the dynamical evolution of the quantum states in the 4d Lorentzian theory~\cite{rovelli2}. In particular, it maps $SU(2)$ states onto $SL(2,\mathbbm{C})$ states. The relation between the interpretations of the Immirzi parameter stemming from the covariant and canonical pictures is non-obvious and remains an open question. With respect to loop quantum cosmology, a model with the Immirzi parameter promoted to be a dynamical field showed that this field serves as a relational time parameter \cite{bombacigno2016}. This corroborates these previous findings of $\beta$ being related to relating the spatial geometry to the `time' direction. One possible development of these ideas could potentially link the gauge connection formalism of loop quantum gravity to Ho\v{r}ava-Lifshitz gravity, which explicitly breaks spacetime diffeomorphism invariance in favor of having an explicit time variable as is in the case of quantum mechanics~\cite{horava2009}. What is pursued in the following is a connection to the theory of shape dynamics.

\section{Shape Dynamics}\label{S:sd}

Shape dynamics is a recent reformulation of gravitation which is conformally invariant. The original ideas of shape dynamics were developed by Barbour~\cite{anderson2003,barbour2004,barbour2011} and the current formulation was developed by Gomes, Gryb, and Mercati~\cite{gryb,gomes,gomes2011,gomes2012,mercati2014}. The motivation for introducing conformal invariance is based on conclusions from York's solution to the initial value problem of general relativity~\cite{york1971,york1972}. The initial value conditions of general relativity on a constant mean curvature slice are that of conformal 3-geometries and a rank-2 symmetric transverse-traceless tensor proportional to the transverse-traceless component of the extrinsic curvature. The dynamics are given by the dynamics of conformal 3-geometries embedded into a 4-dimensional spacetime. Shape dynamics takes this concept a step further by describing the dynamics in terms of conformal 3-geometries alone without any embedding into a larger manifold. 

The construction principle of shape dynamics from geometrodynamics involves the introduction of a scalar `St\"uckelberg field' $\varphi(x)$~\cite{ruegg2004} to the canonical phase space~\cite{gomes2012}. This addition of a St\"uckelberg field is a mechanism for revealing a symmetry in a gauge fixed system and essentially allows for the system to be un-gauge fixed. This new extended phase space is called the Linking Theory~\cite{gomes2012}. For the case where the 3d spatial manifold $\Sigma$ is asymptotically flat, the transformation of the symplectic potential is
\begin{equation}
\begin{gathered}
	\Theta_{\mathrm{GM}} 
	= \int_\Sigma \mathrm{d}^3x \tensor{p}{^{ab}} \mathfrak{d} \tensor{q}{_{ab}}
	\\\downarrow\\
	\Theta_{\mathrm{LT}}
	= \int_\Sigma \mathrm{d}^3x e^{-4\varphi} \tensor{\bar{p}}{^{ab}} \mathfrak{d} e^{4\varphi} \tensor{\bar{q}}{_{ab}}
	= \int_\Sigma \mathrm{d}^3x \left( \tensor{\bar{p}}{^{ab}} \mathfrak{d} \tensor{\bar{q}}{_{ab}} + \pi \mathfrak{d} \varphi \right) .
\end{gathered}
\end{equation}
The introduction of $\pi := 4 \bar{p}^{ab} \bar{q}_{ab}$ as the momentum conjugate to the St\"uckelberg field $\varphi$ generates a new conformal constraint $\bar{\mathcal{C}}$ in the linking theory
\begin{equation}
	\bar{\mathcal{C}}[\rho] = \int_\Sigma \mathrm{d}^3x \rho \left( \pi - 4 \bar{p}^{ab} \bar{q}_{ab} \right)
\label{E:geoconf}
\end{equation}
in addition to the conformally modified Hamiltonian and diffeomorphism constraints $\bar{\mathcal{H}}[N]$ and $\bar{\mathcal{D}}[\vec{N}]$. 
The reduction from the linking theory to general relativity is recovered by imposing the second-class gauge fixing constraint $\varphi(x) \approx 0$. This leads to a phase space which is equivalent to the geometrodynamics phase space $(q_{ab} , p^{ab})$. This gauge fixing requires that
\begin{equation}
	\left\{ \varphi(x) , \bar{\mathcal{C}}[\rho] \right\} = \rho(x) \overset{!}{\approx} 0 ,
\end{equation}
which eliminates the conformal constraint from the constraint algebra. 
In contrast, the theory of shape dynamics is obtained from imposing the constraint $\pi(x) \approx 0$~\cite{gomes2012}. 

A step towards quantizing the resulting theory may be found by following the procedure of loop quantum gravity and reconstructing the formalism in terms of a gauge connection. The symplectic potential and structure of the connection-based linking theory may be constructed as 
\begin{align}
	\Theta_{\mathrm{LT}} &= \int_\Sigma \mathrm{d}^3x \left[ \tensor{\bar{E}}{_i^a} \mathfrak{d} \tensor{\bar{A}}{^i_a} + \pi \mathfrak{d} \varphi \right]
,\label{E:symplecticltpot}\\
	\Omega_{\mathrm{LT}} &= \int_\Sigma \mathrm{d}^3x \left[ \mathfrak{d} \tensor{\bar{A}}{^i_a} \wedge \mathfrak{d} \tensor{\bar{E}}{_i^a} + \mathfrak{d} \varphi \wedge \mathfrak{d} \pi \right]	.
\label{E:symplecticlt}
\end{align}
Just as the Ashtekar-Barbero phase space only differs from the geometrodynamics phase space by a canonical transformation, the linking theory constructed here only differs from that of the usual shape dynamics by a similar canonical transform. Along with this new configuration variable $\varkappa \tensor{\bar{A}}{^i_a} = \tensor{\bar{\Gamma}}{^i_a} + \tensor{\bar{K}}{^i_a}$ comes a Gau{\ss} constraint $\bar{D}_a \tensor{\bar{E}}{_i^a} \equiv \partial_a \tensor{\bar{E}}{_i^a} + \tensor{\epsilon}{_{ij}^k} \tensor{\bar{A}}{^j_a} \tensor{\bar{E}}{_k^a} \approx 0$. This constraint may be constructed from adding the 1$^{\mathrm{st}}$~Cartan structure equation $\partial_a \tensor{\bar{E}}{_i^a} + \tensor{\epsilon}{_{ij}^k} \tensor{\bar{\Gamma}}{^j_a} \tensor{\bar{E}}{_k^a} = 0$ to the rotation constraint $\tensor{\epsilon}{_{ij}^k} \tensor{\bar{K}}{^j_a} \tensor{\bar{E}}{_k^a} \approx 0$, which is needed to constrain the extrinsic curvature tensor $K_{ab}$ to be symmetric. In terms of the connection variable, the constraints of the linking theory are
\begin{subequations}
\begin{align}
	\bar{\mathcal{C}}[\rho]	&=	\frac{1}{\varkappa} \int_\Sigma \mathrm{d}^3x \rho \left( 4 \mathcal{K} + \pi \right) \\
	\bar{\mathcal{G}}[\tilde{\xi}]	&=	\frac{1}{\varkappa} \int_\Sigma \mathrm{d}^3x \xi^i \bar{D}_a \tensor{\bar{E}}{_i^a} \\
	\bar{\mathcal{D}}[\vec{N}]	&=	\frac{1}{\varkappa} \int_\Sigma \mathrm{d}^3x N^a \left( \tensor{\bar{E}}{_i^b} \tensor{\bar{F}}{^i_{ab}} - \tensor{\bar{A}}{^i_a} \bar{D}_c \tensor{\bar{E}}{_i^c} + \pi \partial_a \varphi \right) \\
	\bar{\mathcal{H}}[N]	&=	\frac{1}{2\varkappa} \int_\Sigma \mathrm{d}^3x N \frac{e^{2\varphi}}{\sqrt{\bar{E}}} \tensor{\bar{E}}{_i^a}\tensor{\bar{E}}{_j^b} \Big( \tensor{\epsilon}{^{ij}_k} \tensor{\bar{F}}{^k_{ab}} - 2 (1+e^{-8\varphi}) \tensor{\bar{K}}{^i_{[a}}\tensor{\bar{K}}{^j_{b]}} + 8 e^{-\varphi} \eta^{ij} \nabla_a \nabla_b e^{\varphi} \Big) \label{E:conformalhamiltonian}
\end{align}
\end{subequations}
where $\tensor{\bar{F}}{^k_{ab}} = \varkappa \partial_a \tensor{\bar{A}}{^k_b} - \varkappa \partial_b \tensor{\bar{A}}{^k_a} + \varkappa^2 \tensor{\epsilon}{^k_{ij}} \tensor{\bar{A}}{^i_a} \tensor{\bar{A}}{^j_b}$ is the curvature of the new connection and $\mathcal{K} = \tensor{\bar{K}}{^i_a} \tensor{\bar{E}}{_i^a} = \tensor{{K}}{^i_a} \tensor{{E}}{_i^a}$ is the densitized trace of the extrinsic curvature which is conformally invariant by definition. 
Boundary conditions for the asymptotically flat manifold are chosen such that the 4-metric reduces to Minkowski at spatial infinity (radial coordinate $r\to\infty$). These boundary conditions for the linking theory variables and the Lagrange multipliers are~\cite{gomes,thiemann}
\begin{equation}
\begin{aligned}
	\tensor{\bar{A}}{^i_a} &\to \mathcal{O}(r^{-2})
&	\tensor{\bar{E}}{_i^a} &\to \delta_i^a + \mathcal{O}(r^{-1})
\\	e^{4\varphi} &\to 1 + \mathcal{O}(r^{-1})
&	\pi &\to \mathcal{O}(r^{-2})
\\	N &\to 1 + \mathcal{O}(r^{-1})
&	N^a &\to \mathcal{O}(r^{-1})
&	\rho &\to \mathcal{O}(r^{-1})
\end{aligned}
\label{E:bdrycdns}
\end{equation}
As in the usual shape dynamics case, these boundary conditions are required in order to find a solution to the lapse fixing equation.

Incidentally, the linking theory as written in \eqref{E:symplecticlt} can also be obtained by setting the Immirzi parameter to take the form of the conformal transformation $\beta = e^{4\varphi(x)}$. Explicitly, following the procedure of \eqref{E:betatrans}, 
\begin{subequations}
\begin{align}
	\Theta
	&=	\frac{1}{\varkappa} \int_\Sigma \mathrm{d}^3x \tensor{E}{_i^a} \mathfrak{d} \tensor{K}{^i_a}	\\
	&=	\frac{1}{\varkappa} \int_\Sigma \mathrm{d}^3x e^{4\varphi} \tensor{\bar{E}}{_i^a} \mathfrak{d} e^{-4\varphi} \tensor{\bar{K}}{^i_a}	\\
	&=	\frac{1}{\varkappa} \int_\Sigma \mathrm{d}^3x \left[ \tensor{\bar{E}}{_i^a} \mathfrak{d} \tensor{\bar{K}}{^i_a} - 4\mathcal{K} \mathfrak{d} \varphi \right] + \frac{1}{\varkappa} \int_\Sigma \mathrm{d}^3x \tensor{\bar{E}}{_i^a} \mathfrak{d} \tensor{\bar{\Gamma}}{^i_a}	\\
	&=	\frac{1}{\varkappa} \int_\Sigma \mathrm{d}^3x \left[ \tensor{\bar{E}}{_i^a} \mathfrak{d} \left( \tensor{\bar{\Gamma}}{^i_a} + \tensor{\bar{K}}{^i_a} \right) - 4\mathcal{K} \mathfrak{d} \varphi \right]	\\
	&=	\int_\Sigma \mathrm{d}^3x \left[ \tensor{\bar{E}}{_i^a} \mathfrak{d} \tensor{\bar{A}}{^i_a} + \pi \mathfrak{d} \varphi \right]	,
\end{align}
\end{subequations}
which is precisely the same $\Theta$ as that of the linking theory \eqref{E:symplecticltpot}. No anomalous Immirzi parameter appears since any constant Weyl scaling may simply be absorbed into the conformal factor $\varphi$.

\section{Shape Dynamical Connection Theory}\label{S:connectionsd}
Analogously to the standard linking theory, there are two gauge fixings, $\varphi(x) - \varphi_0 \approx 0$ and $\pi(x) - \pi_0 \approx 0$, which are implemented as second-class constraints. The choice $\varphi_0 = 0$ reduces the conformal variables $( \tensor{\bar{A}}{^i_a} , \tensor{\bar{E}}{_i^a} )$  down to the nonconformal variables $( \tensor[^{(1)}]{\!{A}}{^i_a} , \tensor{{E}}{_i^a} )$, i.e. a phase space which is equivalent to the Ashtekar-Barbero phase space for $\beta = 1$ as $\varkappa\ \tensor[^{(1)}]{\!A}{^i_a} = \tensor{\Gamma}{^i_a} + \tensor{K}{^i_a}$. 
In order to recover the usual $\beta \in GL(1,\mathbbm{R})$ dependent phase space $( \tensor[^{(\beta)}]{\!{A}}{^i_a} , \tensor{{E}}{_i^a} )$, the conformal transform used to build the linking theory needs to be altered. Instead of using $\varphi_0 = 0$ as a constraint parameter, the constraint $\varphi_0 = \frac14 \ln \beta$ should be used. 
However, since $\varphi_0 \neq 0$, it is not obvious how to handle this choice of gauge fixing. The literature so far has only developed gauge fixings of the linking theory for either $\varphi_0 = 0$ or $\pi_0 = 0$. Exploring the possible gauge fixing which reduces to the Ashtekar-Barbero-Immirzi variables could be an interesting route for future research. 
These choices will not be explored here as the goal here is to eliminate the $\beta$ ambiguity from the theory altogether. 

The analog of shape dynamics from the connection linking theory is obtained from the gauge $\pi(x) \approx 0$. Checking the commutation of this constraint with the linking theory's Hamiltonian constraint results in
\begin{equation}
	\left\{ \pi , \bar{\mathcal{H}}[N] \right\}
	=	\bar{\mathcal{H}}[2N] - \Bigg[ 8 \frac{e^{-6\varphi} \tensor{\bar{E}}{_i^a} \tensor{\bar{E}}{_j^b}}{\sqrt{\bar{E}}}\tensor{\bar{K}}{^i_{[a}}\tensor{\bar{K}}{^j_{b]}} - 8 e^{2\varphi} \sqrt{\bar{E}} ( \nabla^c \varphi \nabla_c N + \nabla^2 N ) \Bigg]
\label{E:}
\end{equation}
which is required to vanish due to the gauge fixing. The first term is proportional to the linking theory Hamiltonian constraint. Analogously to the geometrodynamic shape dynamics case, requiring this quantity to vanish gives the Lichnerowicz-York equation in terms of the connection variables. Defining $\Phi(x) := e^{\varphi(x)}$, the solution to this equation is given by $\Phi_o[\tensor{\bar{A}}{^i_{a}}, \tensor{\bar{E}}{_i^a}; x)$.\footnote{The hybrid notation $\phi[J_m;y_n)$ indicates that $\phi$ is a functional of functions $\{J_m\}$ and a function of variables $\{y_n\}$.} The solution $\varphi_o[\tensor{\bar{A}}{^i_{a}}, \tensor{\bar{E}}{_i^a}; x)$ is then obtained from taking $\varphi_o = \ln \Phi_o$.
Demanding that the second term vanish results in the `lapse fixing equation'
\begin{equation}
	\nabla^2 N + \nabla^c \varphi \nabla_c N - \frac{e^{-8\varphi} \tensor{\bar{E}}{_i^a} \tensor{\bar{E}}{_j^b}}{{\bar{E}}}\tensor{\bar{K}}{^i_{[a}}\tensor{\bar{K}}{^j_{b]}} = 0 .
\end{equation}
This is a second-order partial differential equation for the lapse $N$. The solution of this equation is denoted by $N_o[\varphi_o, \tensor{\bar{A}}{^i_{a}}, \tensor{\bar{E}}{_i^a}; x)$ which uses the solution $\varphi_o$ of the Lichnerowicz-York equation.
\begin{figure}[h]
\centering
	\includegraphics[width=0.5\linewidth]{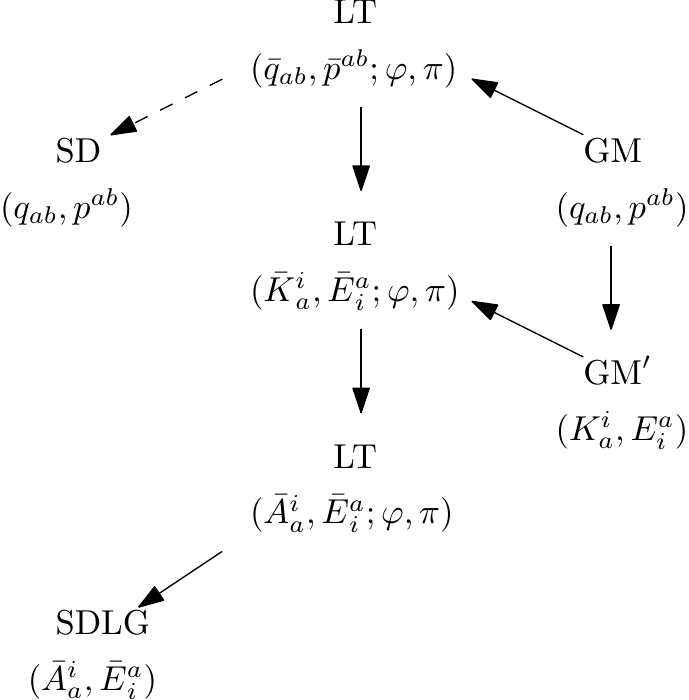}
	\caption{The solid arrows show the canonical transformation and gauge fixing sequence followed to construct the shape dynamical loop gravity (SDLG) theory from geometrodynamics (GM) via a linking theory (LT). Alternatively, the linking theory may also be constructed after making the transformation to geometrodynamics in the triad variables (GM$^\prime$). The dashed line represents the gauge fixing which leads to the conventional shape dynamics (SD) theory.}
	\label{F:flowchart}
\end{figure}
As observed by Refs.~\cite{calcagni2009,menamarugan2002}, the $SU(2)$ connection may only be constructed on the original geometrodynamic manifold if $\beta$ is a constant. 
Instead of privileging the (gauge fixed) geometrodynamics phase space, the theory here privileges the (non-gauge fixed) linking theory. It is in the context of the linking theory that the gauge connection is now constructed from a canonical transform. After the canonical change of phase space to the connection variable in the linking theory, the gauge fixing reduction to shape dynamics is performed. This procedure is sketched diagrammatically in Figure \ref{F:flowchart}.

After reducing the phase space with the gauge fixing, the symplectic structure takes the form
\begin{align}
	\Omega_{\mathrm{SDLG}} 
	&= \int_\Sigma \mathrm{d}^3x \mathfrak{d} \tensor{\bar{A}}{^i_a} \wedge \mathfrak{d} \tensor{\bar{E}}{_i^a} 
\end{align}
where the conformal factor has disappeared from the phase space due to the gauge fixing constraint of $\pi(x) \approx 0$. The consensus in the standard treatment of loop quantum gravity, using $SU(2)$ as the gauge group is considered to be aesthetically undesirable due to the loss of internal boosts as a gauge degree of freedom. Here in the context of shape dynamics, this is no longer an issue since Lorentz symmetry is merely an emergent symmetry rather than a fundamental one~\cite{carlip2015}. Thus the gauge group $SU(2)$ is actually more desirable in the context of shape dynamics. 
The constraints after the gauge fixing are 
\begin{subequations}
\begin{align}
	\bar{\mathcal{G}}[\tilde{\xi}]	&=	\frac{1}{\varkappa} \int_\Sigma \mathrm{d}^3x \xi^i \bar{D}_a \tensor{\bar{E}}{_i^a}	\\
	\bar{\mathcal{D}}[\vec{N}]	&=	\frac{1}{\varkappa} \int_\Sigma \mathrm{d}^3x N^a \left( \tensor{\bar{E}}{_k^b} \tensor{\bar{F}}{^k_{ab}} - \tensor{\bar{A}}{^i_a} \bar{D}_c \tensor{\bar{E}}{_i^c} \right)	\\
	\bar{\mathcal{C}}[\rho]		&=	\frac{4}{\varkappa} \int_\Sigma \mathrm{d}^3x \rho \mathcal{K}
\end{align}
\end{subequations}
The algebra of these constraints reads as 
\begin{equation}
\begin{aligned}
	\left\{ \bar{\mathcal{G}}[\tilde{\xi}] , \bar{\mathcal{G}}[\tilde{\xi}'] \right\}	&=	\bar{\mathcal{G}}\left[[\tilde{\xi} , \tilde{\xi}']\right]	&	\left\{ \bar{\mathcal{G}}[\tilde{\xi}] , \bar{\mathcal{D}}[\vec{N}] \right\}	&=	\bar{\mathcal{G}}\left[\mathsterling_{\vec{N}} \tilde{\xi}\right]	\\
	\left\{ \bar{\mathcal{D}}[\vec{N}] , \bar{\mathcal{D}}[\vec{N}'] \right\}	&=	\bar{\mathcal{D}}\left[ [\vec{N} , \vec{N}'] \right]	&	\left\{ \bar{\mathcal{C}}[\rho] , \bar{\mathcal{D}}[\vec{N}] \right\}	&=	\bar{\mathcal{C}}\left[\mathsterling_{\vec{N}} \rho\right]	\\
	\left\{ \bar{\mathcal{C}}[\rho] , \bar{\mathcal{C}}[\rho'] \right\}	&=	0	&	\left\{ \bar{\mathcal{G}}[\tilde{\xi}] , \bar{\mathcal{C}}[\rho] \right\}	&=	0	,
\end{aligned}
\end{equation}
which is much simpler than the usual loop quantum gravity case due to the lack of a complicated relation of the bracket of two scalar constraints resulting in a vector constraint. Of particular note, there is nothing as complicated as the commutator of two Hamiltonian constraints resulting in a diffeomorphism constraint. This point will be elaborated more on in the following section which discusses the quantum theory. Note that the conformal and Gau{\ss} constraints commuted with the conformal constraint vanish strongly. The Gau{\ss}--conformal bracket is equal to the 1$^{\mathrm{st}}$~Cartan structure equation which does not vanish as a constraint, but rather as an identity. The bracket therefore vanishes strongly. The Poisson commutation of two conformal constraints results in two identical terms with opposing signs, so this too vanishes strongly. 

The total Hamiltonian which generates the dynamics is given by $H_T = H_{\mathrm{gl}}(\varphi_o,N_o) + \bar{\mathcal{D}}[\vec{N}] + \bar{\mathcal{C}}[\rho] + \bar{\mathcal{G}}[\tilde{\xi}]$ where $H_{\mathrm{gl}}$ is the global Hamiltonian.
The standard formulation of shape dynamics considers the asymptotically flat case to be an effective theory which is valid locally in the universe while the `full' theory requires the consideration of a spatially compact manifold and volume preserving conformal transformations. In particular, the unconstrained Weyl transform implemented in the linking theory above is only obtained from deparameterization of this theory with respect to the York time. This deparameterization then results in a global Hamiltonian which generates the dynamics of the theory with respect to a time parameter \cite{mercati2014}. See Appendix~\ref{S:compact} for an outline of the spatially compact manifold case and how the global Hamiltonian is obtained. For the case of an asymptotically flat subsystem, the Hamiltonian is given by the ADM mass \cite{koslowski2015b}.

One point to make here is that the order of the canonical transformations is non-commutative. The above construction invoking the linking theory requires that the conformal transformation is implemented before the gauge connection $\bar{A}$ is constructed. Had the usual Ashtekar-Barbero variables been constructed prior to the conformal transformation, the resulting conformal variables would then be
\begin{subequations}
\begin{align}\label{E:nonconformalconnection}
	\varkappa \tensor{A}{^i_a}	&=	e^{-4\varphi_o} \beta \tensor{\bar{K}}{^i_a} + \tensor{\bar{\Gamma}}{^i_a} + 2\tensor{\epsilon}{^i_j^k} \tensor{\bar{E}}{^j_a} \tensor{\bar{E}}{_k^c} \partial_c \varphi_o	\\
	\tensor{E}{_i^a}	&=	e^{4\varphi_o} \tensor{\bar{E}}{_i^a}	.
\end{align}
\end{subequations}
This different expression for the connection is a result of the transformation $\tensor{{\Gamma}}{^i_a}[E] \to \tensor{{\Gamma}}{^i_a}[e^{4\varphi}\bar{E}] = \tensor{\bar{\Gamma}}{^i_a} + 2\tensor{\epsilon}{^i_j^k} \tensor{\bar{E}}{^j_a} \tensor{\bar{E}}{_k^c} \partial_c \varphi$. As evidenced by the asymmetric scaling of $\tensor{K}{^i_a}$ and $\tensor{\Gamma}{^i_a}$ in~\eqref{E:nonconformalconnection}, these variables are not equivalent to those of the linking theory constructed above where the connection is simply $\varkappa \tensor{\bar{A}}{^i_a} = \tensor{\bar{\Gamma}}{^i_a} + \tensor{\bar{K}}{^i_a}$. The absorption of $\beta$ into the conformal factor is still possible here, but the parameter would still appear explicitly in the last term of~\eqref{E:nonconformalconnection}. Furthermore, the transformation~\eqref{E:nonconformalconnection} does not allow for a Gau{\ss}-type constraint in the conformal variables which then jeopardizes one of the key aspects of the loop quantization program. For this reason, the construction here first forms the linking theory with the triadic geometrodynamic variables, and only then constructs the analog of the Ashtekar-Barbero connection. 

A treatment of the Immirzi parameter as a conformal scale factor in a manner similar the construction of the linking theory here was previously performed by Wang~\cite{wang2005b,wang2008}. However, in that case the conformal superspace (roughly equivalent to the linking theory here) is considered to be the fundamental phase space and it is that space in which the quantization is attempted. Also in the case of Ref.~\cite{wang2005b}, the additional canonical variables of the conformally extended phase space are the York time ($\propto$ trace of the extrinsic curvature) and the volume element ($\sqrt{\bar{q}}$). The case of Ref.~\cite{wang2008} considers the object $\Phi^4$ as a replacement for the Immirzi parameter and $\Phi$ as an additional canonical variable. The constructed theory in that case is similar to the connection based linking theory presented here. 

A potential loop quantization procedure would involve quantizing $\Phi$ the way scalar fields are quantized in loop quantum gravity, either with point holonomies~\cite{chiou,thiemann} or via polymer quantization~\cite{kaminski2006a,kaminski2006b}. A theory built upon this idea was previously explored~\cite{wong2017}. A complication which arises in this theory is that the Hamiltonian constraint contains terms such as $\Phi^{4}$, $\Phi^{-4}$, and $\Phi \nabla^2 \Phi$. While it is technically possible to promote such terms to quantum operators~\cite{zhang2011a,zhang2011b}, their form is extremely complicated and a solution to this constraint operator would be even more challenging than the corresponding operator in the standard loop quantum gravity formalism. It is therefore questionable what the advantage of quantizing such a theory would be. One could postulate that building spin network states which manifestly incorporate conformal symmetry would force the Hamiltonian constraint operator to have a simple action, but it remains far from clear how this idea would be implemented in practice. The shape dynamical theory explored in this article has a much cleaner formulation and thus could be argued to possesses stronger motivation for its construction (assuming the standard construction of shape dynamics is believed to be well-motivated).

A previous entry into the literature also considered a quantization scheme of general relativity with an eliminated Hamiltonian constraint and an added conformal constraint~\cite{bodendorfer2013a,bodendorfer2013b}. The theory there is constructed by conformally coupling general relativity with a dynamical scalar field. The quantization procedure is also based upon the framework of loop quantum gravity, but the Immirzi parameter remains explicit and no conformal interpretation of the parameter is acknowledged. 

An interesting parallel with standard loop quantum gravity can be made with the conformal constraint. Instead of constructing the linking theory from the conformal scaling, choosing $\varphi(x) = \varphi_o = \mathrm{const.} = \frac14\ln\beta$ leads to the standard formulation of the Ashtekar-Barbero variables. As mentioned previously, there exist some desirable features for demanding that $\beta = i$. A method for recovering this value from the case $\beta = 1$ is by using a phase space `Wick rotation' whose generator is
\begin{equation}\label{E:wick}
	C_W = \frac{1}{\varkappa} \int_\Sigma \mathrm{d}^3x \tensor{K}{^i_a} \tensor{E}{_i^a} .
\end{equation}
This is the generator of an `analytic continuation' in the {\em phase space} in contrast to the conventional Wick rotation which maps the coordinate $t \to -i t$. The complex connection is derived from the chain of transformations
\begin{equation}
	( A = \Gamma + K , E ) \to ( K , E ) \to ( i K , -i E ) \to ( \tensor[^{\mathbb{C}}]{A}{} = \Gamma + i K , \tensor[^{\mathbb{C}}]{E}{} = -i E ) .
\end{equation}
The transformation to the third set is obtained from
\begin{align}
	i K	&=	\sum_{n=0}^{\infty} \left(\frac{i \pi}{2}\right)^n \frac{\{ C_W , K \}^{(n)}}{n!}	\\
	-i E	&=	\sum_{n=0}^{\infty} \left(\frac{i \pi}{2}\right)^n \frac{\{ C_W , E \}^{(n)}}{n!}
\end{align}
with $\{ C_W , f \}^{(n+1)} := \{ C_W , \{ C_W , f \}^{(n)} \}$ and $\{ C_W , f \}^{(0)} := \{ C_W , f \}$. Note that the generator \eqref{E:wick} is proportional to $\bar{\mathcal{C}}[1]$ above. The transformation for the case considered by Barbero~\cite{barbero1995} and Immirzi~\cite{immirzi1997a,immirzi1997b} may be obtained from the replacement\footnote{This replacement works by noting that $i = e^{i\pi/2}$ gives the replacement $i\pi/2 \to \ln\beta$ in the sum.} $i \to \beta$ as
\begin{align}
	\sum_{n=0}^{\infty} \left( \ln\beta \right)^n \frac{\{ C_W , K \}^{(n)}}{n!}	&= \beta K , \\
	\sum_{n=0}^{\infty} \left( \ln\beta \right)^n \frac{\{ C_W , E \}^{(n)}}{n!} &= \frac{E}{\beta} .
\end{align}
The phase space-Wick rotation then, instead of mapping to a complex phase space, maps the variables to a conformally scaled phase space. This is further evidence that the ad hoc replacement $\varkappa A = \Gamma + i K \to \Gamma + \beta K$ originally implemented by Barbero~\cite{barbero1995} and Immirzi~\cite{immirzi1997a,immirzi1997b} does indeed allow for the interpretation that $\beta$ represents a rescaling of the geometry. The original goal of this phase space-Wick rotation was to somehow recover the original $SL(2,\mathbbm{C})$ degrees of freedom which were lost from choosing $\beta \in GL(1,\mathbbm{R})$. This calculation implies that by using a real parameter in the phase space-Wick rotation instead of an imaginary one, the rotation adds a conformal degree of freedom rather than a boost degree of freedom. This is similar in spirt to the construction principles of shape dynamics. 

\section{Quantum Shape Dynamical Loop Gravity}\label{S:qsdlg}

Now that the variables of the theory are in terms of a gauge connection and a densitized triad, the same loop quantization procedure utilized in loop quantum gravity may now be applied to the theory resulting from the $\pi(x) \approx 0$ gauge fixing of the linking theory. This involves constructing $SU(2)$ holonomies $\bar{U}$ from the connection 1-form $\varkappa \bar{\boldsymbol{A}}$ and fluxes $\bar{\mathcal{E}}$ from the 2-form dual of the triad vector density $*\bar{\boldsymbol{E}} = \tensor{\epsilon}{_{abc}} \tensor{\bar{E}}{_{i}^a} \tau^{*i} \otimes \mathrm{d}x^b \wedge \mathrm{d}x^c$. 
The quantization procedure at the kinematical level follows the quantization procedure of standard loop quantum gravity~\cite{cianfrani,chiou,kiefer,sahlmann2012,thiemann}. As in the standard formulation, the kinematical quantum states are s-knots, which by construction solve the Gau{\ss} and diffeomorphism constraints. 

Since the kinematical quantization procedure is mathematically equivalent to the standard formulation, it is still possible to construct geometric operators for this new shape dynamics-like theory. The reconstructed area operator has eigenvalues given by
\begin{equation}
	\hat{\bar{\mathcal{A}}}_\sigma | \Psi \rangle = \ell_p^2 \sum_{l\in\sigma} \sqrt{j_l (j_l + 1)} | \Psi \rangle .
\end{equation}
In contrast with the standard formalism, there is no anomalous parameter rescaling the Planck length. Quantizing in the conformal variables is therefore non-ambiguous about the fundamental length scale: it is the ordinary Planck length. The physical (non-conformal) geometry still involves a rescaling (from $e^{4\varphi_o}$), in the conformal frame given by the geometry of $\bar{q} \bar{q}_{ab} = \eta_{ij} \tensor{\bar{E}}{^i_a} \tensor{\bar{E}}{^j_b}$ there is no ambiguity surrounding the fundamental length scale.

The Gau{\ss} and diffeomorphism constraints may be implemented as they are in the standard loop quantum gravity framework. Like the Hamiltonian constraint in standard loop quantum gravity, here the conformal constraint is implemented as an operator. The mathematical treatment of this operator makes use of the same techniques used to quantize the Hamiltonian constraint of the usual theory. The relevant quantity which needs to be transformed into a quantum operator is the densitized trace of the extrinsic curvature. This term actually appears explicitly in the Hamiltonian constraint of the standard theory, so all of the mathematical formalism developed there is directly applicable to the conformal constraint here. It therefore follows immediately that the quantum conformal constraint operator will not necessitate any additional technical difficulties to what has already been accomplished in the standard formalism. The quantum conformal constraint operator borrows many desirable characteristics from the standard quantum Hamiltonian constraint operator such as $SU(2)$ invariance and cylindrical consistency. It is worth emphasizing that this cylindrical consistency is due to the fact that the variables, particularly the connection, are still the conformally transformed ones. A problem encountered with trying to use the non-conformal variables has to do with the way the connection transforms, cf. \eqref{E:nonconformalconnection}. The complicated behavior of the connection under conformal transformations would end up breaking the cylindrical consistency of the conformal constraint thereby preventing an acceptable quantization in terms of the holonomies.  

The conversion of the conformal constraint into a quantum operator is done in a procedure analogous to that of the Hamiltonian constraint in the standard formulation of loop quantum gravity. As is the case with the standard formulation of loop quantum gravity, a proper mathematically rigorous treatment of this operator is as-of-yet unobtainable. What follows should only be interpreted as a rough sketch of what the quantum conformal constraint operator's properties might be. 

Making use of certain Poisson brackets (a technique known as ``Thiemann's trick'' in the literature), the conformal constraint can be rewritten as
\begin{equation}
	\bar{\mathcal{C}}[\rho] = \frac4\varkappa \int_\Sigma \mathrm{d}^3x \rho \mathcal{K} = \frac4\varkappa \int_\Sigma \mathrm{d}^3x \rho \left\{ \bar{\mathcal{F}} , \bar{V}(R) \right\}
\end{equation}
where
\begin{align}\label{E:curvtrace}
	\bar{\mathcal{F}} &:= \frac{\tensor{\bar{E}}{_i^a} \tensor{\bar{E}}{_j^b}}{\sqrt{\bar{E}}} \tensor{\epsilon}{^{ij}_k} \tensor{\bar{F}}{^k_{ab}} = *\mathrm{Tr}\left( \bar{\boldsymbol{F}} \wedge \{ \bar{\boldsymbol{A}} , \bar{V} \} \right)
\end{align}
and
\begin{align}
	\bar{V}(R) &= \int_R \mathrm{d}^3x \sqrt{\bar{E}} .
\end{align}
The object ${\bar{\mathcal{F}}}$ used here is of the same form as what is called the ``Euclidean Hamiltonian constraint'' in the standard loop quantum gravity literature. 
As in the standard loop quantum gravity framework, a suitable regularization needs to be imposed. A regularization allows for \eqref{E:curvtrace} to be rewritten in terms of holonomies along the edges of a lattice rather than the connection. Following the regularization scheme proposed in Ref.~\cite{thiemanni}, the quantum conformal constraint operator may then be written as
\begin{equation}
	\hat{\bar{\mathcal{C}}}[\rho] = -\frac{8}{3\varkappa\hbar^2} \sum_{v\in\Delta} \rho(v) \left[ \mathrm{Tr} \left( \hat{\bar{U}}_{\alpha_{[ij]}(\Delta)} \hat{\bar{U}}_{s_k(\Delta)} \left[ \hat{\bar{U}}^{-1}_{s_k(\Delta)} , \hat{\bar{V}}(R) \right] \right) , \hat{\bar{V}}(R) \right]
\label{E:confop}
\end{equation} 
where the sum is over vertices $v$ of a spin network with regulator $\Delta$. The labelings $\alpha_{[ij]}(\Delta)$ and $s_k(\Delta)$ label particular edges of the regulating lattice $\Delta$. See e.g. Refs.~\cite{alesci2013,thiemanni} for details. This quantum operator is of the same form as the quantum operator corresponding to $\mathcal{K}$, which already exists in the standard loop quantum gravity literature as it is an integral component of the ``Lorentzian'' Hamiltonian constraint operator. The explicit action of the extrinsic curvature operator on an example trivalent node is presented in Ref.~\cite{alesci2013}. The explicit matrix elements of this operator take a computationally complicated form, but heuristically are proportional to a sum of matrix elements of the volume operator. In order for the action of the conformal constraint operator to annihilate a quantum state, it could be imagined that there exists some equivalence class condition on allowed spin networks such that the matrix elements of the volume operator induced by the action of \eqref{E:confop} conspire to annihilate the state. There does not, however, currently exist an explicit form for the matrix elements of the volume operator as there does, for instance, the area operator. This impedes on the ability to produce a well-defined criterion on the spin network nodes such that states are annihilated by \eqref{E:confop}. 

Due to the construction of the quantum conformal constraint operator, it shares some of the same technical problems as the Hamiltonian constraint operator of standard loop quantum gravity, with one specific ambiguity being the choice of regularization. The discussion here is based upon the regularization scheme proposed in Ref.~\cite{thiemanni}. The operator \eqref{E:confop} behaves differently under different regularization schemes and therefore understanding how it may annihilate quantum states is regularization dependent. 
Another criterion which needs to be fulfilled is
\begin{equation}
	\left[ \hat{\bar{\mathcal{C}}}[\rho] , \hat{\bar{\mathcal{C}}}[\rho'] \right] |\Psi\rangle \overset{!}{=} \hat{0} |\Psi\rangle
\label{E:ccalg}
\end{equation}
which is needed to satisfy the quantum analog of the classical constraint algebra. Given that $\hat{\bar{\mathcal{C}}}$ is similar in form to the Euclidean Hamiltonian constraint of conventional loop quantum gravity, \eqref{E:ccalg} should hold on-shell. However, also analogously to the conventional case, it remains to be shown that this relation is anomaly-free off-shell.

\section{Conceptual Compatibility}\label{S:cc}

While there are no obvious technical difficulties in the above construction (beyond those of the standard loop quantum gravity formalism), it is not completely obvious that the conceptual principles of loop quantum gravity are exactly compatible with those of shape dynamics. 
Shape dynamics is a conformally invariant theory. Therefore there should not be any fixed geometric quantities. For example, any nonzero area $\mathcal{A}$ is conformally equivalent to any other nonzero area $\mathcal{A}'$. Loop quantum gravity on the other hand, has specific values for the quanta of area: those which are the eigenvalues of the area operator. A caveat here is that the areas measured by the area operator are not Dirac observables in loop quantum gravity. 
A previous discussion on the conceptual reconciliation between loop quantum gravity and shape dynamics~\cite{smolin2014} suggested that instead of area and volume operators, the key geometric operator should be an angle operator as angles are invariant under conformal transformations. 

The `quantum shape dynamical loop gravity' constructed here is somewhat different than the loop quantization of shape dynamics heuristically outlined previously in Refs.~\cite{koslowski2013,koslowski2015}. 
The approach there is based on the standard (non-conformal) loop variables, unlike the treatment with the transformed variables presented here. The connection $\varkappa {A} = {\Gamma} + \beta {K}$ with the non-conformally interpreted $\beta$ is constructed prior to the gauge unfixing to the linking theory. This is the opposite order of the construction of theory described above where the linking theory is obtained first and the connection built afterwards {\em inside} the linking theory. 
In the case of Ref.~\cite{koslowski2013}, the conformal constraint is not promoted to an operator, but it is solved by imposing certain equivalence class conditions on the allowed spin network functions. The underlying idea behind this is to implement what is physically required for spatial conformal invariance. It turns out that this method is at least partially compatible with what has been outlined in the theory here. From the action of the quantum conformal constraint operator~\eqref{E:confop}, it is implied that two spin network states should have volume matrix elements satisfying certain criteria. These criteria could be related to one of the equivalence class conditions imposed in Ref.~\cite{koslowski2013}, wherein any non-zero volume must be conformally equivalent to any other non-zero volume. Another equivalence required is that areas are also conformally invariant. In the theory constructed here, there is no explicit demand that area eigenvalues are conformally related like there is for volume eigenvalues. This equivalence demand could presumably be worked in by hand, but it would be better to show that it is a consequence of some aspect of the theory.

In closing, this paper presents a novel construction of a theory which is build from combining the construction procedures of both loop quantum gravity and shape dynamics. The construction hinges on the observation that the Immirzi parameter in the classical setup of loop quantum gravity arises from a conformal transformation of the geometrodynamic variables. This observation is exploited to construct both the gauge connection variable of the Ashtekar-Barbero formalism and the linking theory of shape dynamics. The resulting shape dynamical loop gravity is free from any ambiguous Immirzi parameter.

\begin{acknowledgments}
This work was supported in part by the DFG through GSC 260.
\end{acknowledgments}

\appendix
\section{Spatially Compact Manifold}\label{S:compact}
The construction of the theory presented above was performed under the assumption that $\Sigma$ was asymptotically flat. This was done to make a clear relationship between $\beta$ and $\varphi$. It is, however, possible to also consider the case where $\Sigma$ is a compact manifold without boundary~\cite{gomes,gomes2011,gomes2012,gryb}. To construct shape dynamics from general relativity in the compact manifold case, volume preserving Weyl transformations are required in place of the unconstrained transformations utilized above. Under a volume preserving conformal transformation, the 3-metric scales as
\begin{equation}
	\mathcal{T}_{\hat{\varphi}} q_{ab} = e^{4\hat{\varphi}} \bar{q}_{ab}
\end{equation}
where the volume preserving conformal factor is given by
\begin{equation}
	\hat{\varphi} = \varphi - \frac16 \ln \langle e^{6\varphi} \rangle
\end{equation}
and $\langle f(x) \rangle = \frac{\int \mathrm{d}^3x \sqrt{q} f(x)}{\int \mathrm{d}^3x \sqrt{q}}$. The scaling for the conjugate momentum is more complicated as
\begin{equation}
	\mathcal{T}_{\hat{\varphi}} p^{ab} = e^{-4\hat{\varphi}} \left( \bar{p}^{ab} - \tfrac13 \sqrt{\bar{q}} \langle \bar{p} \rangle (1 - e^{6\hat{\varphi}}) \bar{q}^{ab} \right) .
\end{equation}
The conformal constraint also takes the more complicated form of
\begin{equation}
	\bar{\mathcal{C}}[\rho] = \int_\Sigma \mathrm{d}^3x \rho \left[ \hat{\pi} - 4( \bar{p} - \sqrt{\bar{q}} \langle \bar{p} \rangle ) \right]
\end{equation}
where $\hat{\pi}$ is the momentum conjugate to $\hat{\varphi}$. 
This formalism requires the use of a volume constraint
\begin{equation}
	\int_\Sigma \mathrm{d}^3x \sqrt{q} \left( e^{6\varphi_o[q,p;x)} - 1 \right) \approx 0
	\label{E:volconstraint}
\end{equation}
where $e^{\varphi_o[q,p;x)}$ is the solution of the Lichnerowicz-York equation. 

Converting from this volume preserving conformal transformation to an unconstrained conformal transformation may be done via a deparameterization with respect to the York time $\tau = \frac23 \langle \bar{p} \rangle$~\cite{barbour2014,mercati2014}. This deparametrization leaves the York time as an internal clock in the classical theory. The conjugate momentum to $\tau$ may be identified as the volume $p_\tau=V=\int \mathrm{d}^3x \sqrt{q}$, since $\{\tau,V\}=1$. The volume constraint~\eqref{E:volconstraint} then takes the form of a time reparameterization (`super-Hamiltonian') constraint $H(t) - p_t \approx 0$. 

This deparameterization results in the shift $\hat{\varphi} \to \varphi$ and the simple transformation for the momentum $\mathcal{T}_{{\varphi}} p^{ab} = e^{-4{\varphi}} \bar{p}^{ab}$. From this stage the above canonical transformation to connection variables is possible with an Immirzi parameter absorbed into the conformal factor. The formalism then proceeds as performed in the main body of this paper, with the addition of a global Hamiltonian
\begin{equation}
	H_{\text{gl}}(\tau) = \int_\Sigma \mathrm{d}^3x \sqrt{q} e^{6\varphi_o[q,p,\tau;x)}
\end{equation}
which generates dynamical evolution with respect to the York time $\tau$. This theory is fully invariant under conformal transformations. 

Generally, the quantization of a super-Hamiltonian constraint results in an evolution equation of the form $\hat{H}_S \psi = 0 \leftrightarrow \hat{H}(t) \psi = \hat{p}_t \psi$. This implies that in the quantum theory there exists a Schr{\"o}dinger equation of the form
\begin{equation}
	\hat{H}_{\text{gl}} |\Psi\rangle = \frac{\hbar}{i} \frac{\delta}{\delta\tau} |\Psi\rangle
\end{equation}
or, equivalently, 
\begin{equation}
	\hat{H}_{\text{gl}} |\Psi\rangle = \hat{V} |\Psi\rangle
\end{equation}
which generates the dynamics of the theory. 

\bibliography{qsdlg}

\end{document}